\RequirePackage{fixltx2e}
\documentclass[twocolumn, aps, prl, superscriptaddress, floatfix]{revtex4-1}

\usepackage[colorlinks, breaklinks]{hyperref}
\usepackage{lmodern}
\usepackage{amsfonts, amssymb, amsmath}
\usepackage{graphicx}
\usepackage{color}
\usepackage[utf8]{inputenc}

\newcommand{\uu}{\mathbf{u}}

\newcommand{\ww}{\mathbf{w}}

\newcommand{\xx}{\mathbf{x}}

\newcommand{\ff}{\mathbf{f}}

\graphicspath{{./figs/}}

\begin{document}
	
	\title{Synchronization of Chaos in Fully-Developed Turbulence}

	\author{Cristian C Lalescu}
	\email{clalesc1@jhu.edu}
	\affiliation{Department of Applied Mathematics \& Statistics, The Johns Hopkins University, Baltimore, MD 21218, USA}
	
	\author{Charles Meneveau}
	\email{meneveau@jhu.edu}
	\affiliation{Department of Mechanical Engineering, The Johns Hopkins University, Baltimore, MD 21218, USA}
	
	\author{Gregory L Eyink}
	\email{eyink@jhu.edu}
	\affiliation{Department of Applied Mathematics \& Statistics, The Johns Hopkins University, Baltimore, MD 21218, USA}
	\affiliation{Department of Mechanical Engineering, The Johns Hopkins University, Baltimore, MD 21218, USA}

	\begin{abstract}
		We investigate chaos synchronization of small-scale motions in the three-dimensional turbulent energy cascade, via pseudo-spectral simulations of the incompressible Navier-Stokes equations.
		The modes of the turbulent velocity field below about 20 Kolmogorov dissipation lengths are found to be slaved to the chaotic dynamics of larger-scale modes.
		The dynamics of all dissipation-range modes can be recovered to full numerical precision by solving small-scale dynamical equations with the given large-scale solution as an input, regardless of initial condition.
		The synchronization rate exponent scales with the Kolmogorov dissipation time-scale, with possible weak corrections due to intermittency.
		Our results suggest that all sub-Kolmogorov length modes should be fully recoverable from numerical simulations with standard, Kolmogorov-length grid resolutions.
	\end{abstract}

	\maketitle

		Chaos synchronization (CS) \cite{pecora_carroll_1990} is an intriguing phenomenon which has been defined as ``a process wherein 
		two (or many) chaotic systems \ldots adjust a given property of their motion to a common behavior due to a coupling or to a forcing'' 
		\cite{boccaletti_review}. The simplest example is a chaotic dynamics $\dot{\xx}=\ff(\xx)$ whose phase vector $\xx$ is projected onto 
		two orthogonal components $\xx_1=P_1\xx$ and $\xx_1'=Q_1\xx=\xx-\xx_1$ satisfying two coupled equations
		\begin{equation}
			\begin{aligned}
				\tfrac{d}{dt} \xx_1&= P_1 \ff(\xx_1 + \xx_1'), \\
				\tfrac{d}{dt} \xx_1' &=  Q_1\ff(\xx_1 + \xx_1').
			\end{aligned}\label{original}
		\end{equation}
		Chaos implies sensitive dependence to initial data, with nearby trajectories diverging exponentially. 
	         However, consider another dynamical system in the $Q_1$-space given by a copy of the second equation:
		\begin{equation}
			\frac{d}{dt} \ww= Q_1\ff(\xx_1(t)+ \ww)
			\label{eq:slaved subdynamics}
		\end{equation}
                  with $\xx_1(t)$ substituted from the solution of (\ref{original}). \emph{Chaos synchronization} occurs if the trajectories 
                  $\ww(t)$ and $\xx_1'(t)$ converge, 
                  $\lim_{t\rightarrow\infty}\|\ww(t)-\xx_1'(t)\|=0,$ for an arbitrary choice of initial condition $\ww_0$ in (\ref{eq:slaved subdynamics}). 
                  Such a phenomenon requires that the leading Lyapunov exponent for the subdynamics (\ref{eq:slaved subdynamics}) be 
                  negative. It is often the case that synchronization occurs, at least approximately, even when imperfect data $\tilde{\xx}_1(t)$ is employed
                   in (\ref{eq:slaved subdynamics}), e.g. the exact $\xx_1(t)$ contaminated with substantial errors. 
                   This effect was proposed in \cite{cuomo_etal_1993, xiao_etal_1996, argyris_etal_2005} as a basis for encrypted communications. 
                   CS has also been observed in neural networks 
                   \cite{schiff_etal_1996, chen_etal_2004}, with ``hyper-synchronous'' dynamics in the human brain associated 
                   to epileptic seizures \cite{stam_review_2005}. CS has been reported in spatio-temporal chaos \cite{winful_rahman_1990, 
                   boccaletti_review, kocarev_etal_prl_1997, kocarev_etal_chaos_1997}, 
	          investigated for fluid models used in meteorology \cite{duane_tribbia_2001, duane_etal_2007, duane_oluseyi_2008}, and 
	          proposed as a mechanism for turbulence control \cite{patnaik_wei_2002, guan_etal_2004, boccaletti_bragard_2006}. 

                   No previous numerical study of CS has been made, to our knowledge, for fully-developed three-dimensional Navier-Stokes (NS) turbulence with a 
                   Kolmogorov inertial range. Our goal in this Letter is to explore CS for 3D NS with $P_1$ taken to be the projection onto the finite number of velocity modes 
                   with wavenumber magnitudes less than a fraction $f$ of $\pi/\eta_K,$ where $\eta_K$ is the Kolmogorov disspation scale, and with 
                   $Q_1$ the orthogonal projection onto the modes with higher wavenumbers.  Our principal motivation is experimental results 
                    \cite{anselmet_etal_1984}, theoretical work \cite{paladin_vulpiani_1987,yakhot_sreeni_2005}, and numerical simulations 
                    \cite{schumacher_njp_2007, schumacher_epl_2007}  implying that spatial intermittency can lead to length scales far smaller than the Kolmogorov scale $\eta_K$.
					It has been argued on the basis of such tiny unresolved length scales ``that the DNS [direct numerical simulation] based on the mesh equal to the Kolmogorov scale becomes quite 
		inaccurate'' \cite{yakhot_sreeni_2005}. If true, this would call into question the vast majority of current DNS studies of turbulent flow. 
		A contrary argument  is based on the idea that the sub-Kolmogorov scales should be ``slaved'' to the inertial-range modes and, thus,
		implicit and recoverable from DNS with grid resolution $\eta_K.$ A mathematical formalization of this idea closely related to CS is the notion 
		of an {\it inertial manifold} (IM) \cite{temam_1990}, which consists of an invariant, attractive manifold given by the graph of a mapping $\xx_1'=\Phi(\xx_1)$  
		which recovers $\xx_1'$ for given $\xx_1.$ Existence of an IM with the property of ``asymptotic completeness'' \cite{robinson_1996}
		is one possible mechanism for CS (e.g. see \cite{xieetal_2007}). There are currently no proofs of existence of an IM for 3D NS 
		dynamics, although ``approximate IM'' have been obtained for 2D NS \cite{titi_1990, foias_etal_1993}. These have been proposed 
		for use as nonlinear Galerkin approximations to the dynamics of ``large'' super-Kolmogorov scales in NS turbulence,
		whereas our goal is the opposite one to recover the sub-Kolmogorov scales and address the outstanding issue of the smallest 
		length-scale in a turbulent flow  \cite{schumacher_njp_2007, schumacher_epl_2007}. Even when existence of an approximate IM 
		can be established for NS,  there are no sharp estimates of the smallest fraction $f$ of the Kolmogorov wavenumber  sufficient for 
		slaving. Thus our numerical investigation in this Letter is an important complement to existing mathematical results.   
                   
		The incompressible NS equations with a solenoidal body force $\ff$ have the form:
		\begin{equation}
			\partial_t \uu + P(\uu \cdot \nabla \uu -\nu \Delta \uu)= \ff, 
		\end{equation}
		where $\uu$ is the fluid velocity, $\nu$ is the kinematic viscosity, and $P$ is the Leray projection to enforce the incompressibility condition $\nabla \cdot \uu = 0.$
		We employ the pseudospectral DNS method, which solves a Galerkin approximation to this system 
		\begin{equation}
			\partial_t \uu_2 +P_2(\uu_2 \cdot \nabla \uu_2 - \nu \Delta \uu_2 )= \ff, \label{u2eq}
		\end{equation}
		with $P_2$ the Leray projection in a space spanned by a finite set $B_2$ of Fourier modes. For our purpose, this will represent the  
		``fine-grained'' NS dynamics. In order to study the possible presence of CS,  we consider a further subset $B_1 \subset B_2$ and 
		corresponding projections $P_1$ and $Q_1=P_2-P_1.$
		The subdynamics for the evolution of $\ww \equiv \uu_2 - \uu_1$ is given by the equation:
		\begin{equation}
			\partial_t \ww + Q_1[ (\uu_1+\ww) \cdot \nabla (\uu_1 + \ww) -  \nu \Delta \ww]={\mathbf 0}, 
			\label{eq:refinement field}
		\end{equation}
		where we have assumed that $\ff$ forces only the large scales, i.e. $Q_1\ff ={\mathbf 0}$. In our experiments 
		we shall solve the fine-grained equation (\ref{u2eq}) for $\uu_2(t)$ and then solve the subdynamics (\ref{eq:refinement field})
		with $\uu_1(t)=P_1\uu_2(t).$ We shall investigate whether $\ww(t)$ converges to $Q_1\uu_2(t)$ for increasing $t,$
		independent of the initial data $\ww_0.$ Specifically, we will study the evolution of the normalized error defined as
                  \begin{equation}
			\epsilon(t) = \frac{\|\ww(t) - Q_1\uu_2(t)\|_2}{\|\ww(t)\|_2},
		\end{equation}
		where $\|\|_2$ is the usual $L^2$ norm.
If synchronization occurs,  this error should tend to zero exponentially fast, independent of the initial value $\ww_0.$ 

\begin{figure}[t]
			\includegraphics[width=\columnwidth]{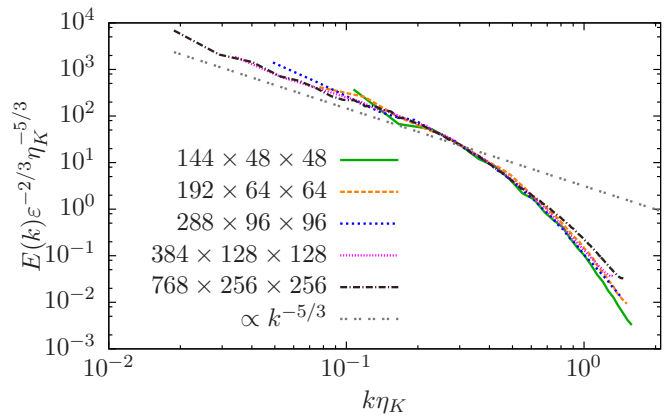}
			\caption{Energy spectra of the Kolmogorov flow simulations for $n=48,64,96,128,256$.
				The spectra are taken from instantaneous snapshots with no time-averaging.
				The moderate quality of the collapse in dissipation-scale units is likely due mostly to strong 
				unsteadiness in the flow.}
			\label{fig:spectra}
		\end{figure}

		The concrete system considered in this work is Kolmogorov flow with $\ff = (A \sin (k_f y), 0, 0)$ for $A = 1$ and $k_f = 1$,
		in an elongated box $[0,L_x]\times [0, L_y]\times [0, L_z]$ with $L_x = 3L_y = 3L_z = 6 \pi$. The numerical simulation uses 
		a space grid of $N = 3n \times n \times n$ points with isotropic mesh-spacing corresponding to maximum wavenumber $k_M=n/2.$ 
		This particular configuration leads to a nontrivial 
		turbulent flow, that is anisotropic and inhomogeneous in the large scales \cite{NS_KFlow_2007}. 		
		In the context of synchronization of chaos, it is relevant that strong bursts can be observed in Kolmogorov flow.
		In \cite{NS_KFlow_2007} very long integration times were used precisely because the time averages presented converge very slowly.
		Thus by studying the system at different times, significantly different regimes can be sampled.
	         In terms of the kinetic energy $E$ and energy dissipation rate $\varepsilon$,
		the Kolmogorov units and the Reynolds number are
$			\eta_K = \left(\frac{\nu^3 }{\varepsilon} \right)^{1/4},\ 
			\tau_K = \left(\frac{\nu   }{\varepsilon} \right)^{1/2},\ 
			R_\lambda = \sqrt{\frac{5}{3}} \frac{2E}{(\nu \varepsilon)^{1/2}}
$			
				Five series of simulations are performed, with resolutions ranging from $144 \times 48 \times 48$ to 
				$768 \times 256 \times 256$ grid points, and $R_\lambda$ going from $40$ up to $250$, keeping the  
				minimum $k_M \eta_K$ around $1.5$.
                  The energy spectra plotted
		in Fig.~\ref{fig:spectra} show a short Kolmogorov inertial range with approximate $-5/3$ power-law scaling. 

\begin{figure}[t]
			\includegraphics[width=\columnwidth]{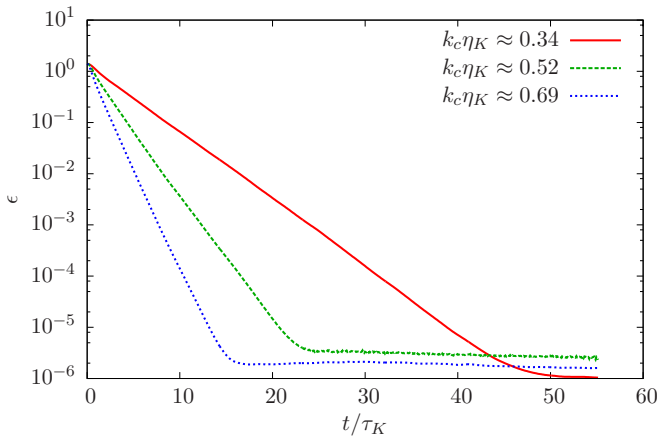}
			\caption{Time evolution of normalized error $\epsilon(t)$ for the simulation on a grid of $308 \times 96 \times 96$ and $R_\lambda \approx 108$, using several cutoff-wavenumbers.
The slope in these graphs yields the exponential decay rate $a,$ or rate of synchonization.
			\label{fig:error decay}}
		\end{figure}

		For our CS study a very long simulation of Kolmogorov flow is performed for each resolution, saving a few time series of the velocity fields 
		from the quasi-stationary regime, each interval separated by relatively long times.
		Four time intervals of $\uu_2$ are chosen for each resolution (three for the $768 \times 256 \times 256$ case). Next
		$\uu_1$ is obtained by the projection $P_1$ of $\uu_2$ onto modes with wavenumbers smaller than a cutoff value in each direction (i.e. $|k_x|, |k_y|, |k_z| < k_c$).
		Finally, $\ww$ is evolved in time using \eqref{eq:refinement field}.
		For each interval, two initial conditions $\ww_0$ were chosen, so that each series consists of eight individual runs.
		In the experiments presented, initial data $\ww_0$ with ``natural'' spectral scaling properties were created 
		by applying random phase shifts to all Fourier modes of $\uu_2 - \uu_1$. 
                   Several alternative initialization methods for $\ww_0$ were tested and yielded consistent results, not shown here.
                   As observed in Fig.~\ref{fig:error decay},  for the indicated values of $k_c$, $\epsilon(t)$ does indeed decrease exponentially fast, 
                   until it reaches a smallest possible value dictated by our single precision arithmetic. Thus $\ww$  synchronizes to $Q_1\uu_2$.

		\begin{figure}[b]
			\includegraphics[width=\columnwidth]{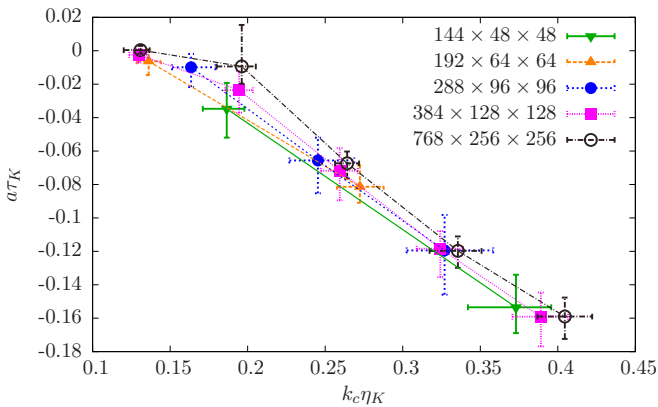}
			\caption{Symbols: average of measured synchronization exponents (obtained by fitting the exponential range in results such as in Fig.~\ref{fig:error decay}) as 
			function of cutoff-wavenumber for five different simulation sizes and Reynolds numbers, 
			plotted in Kolmogorov  units.
Error bars are for maximum and minimum values of different runs and ranges used in the fit.
			\label{fig:exponent stats}}
		\end{figure}

		\begin{figure}[h!]
			\includegraphics[width=\columnwidth]{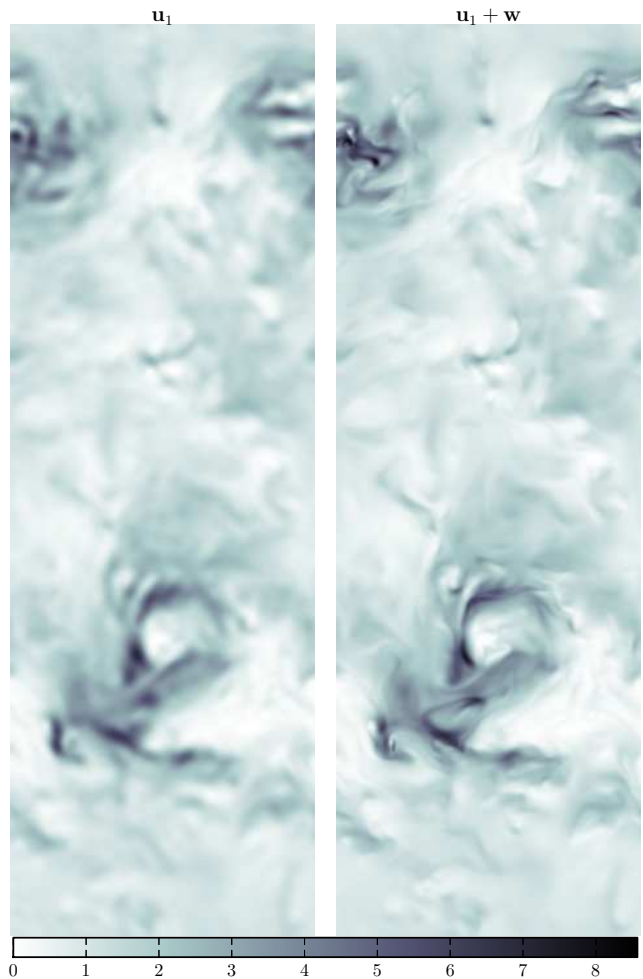}
			\caption{Kinetic energies, taken for a fixed $z$ from a $768 \times 256 \times 256$ simulation ($x$ varies on the vertical and $y$ on the horizontal).
			Left: coarse grained field obtained with $k_c \eta_K \approx 1/4$.
			Right: refined version of coarse grained field.
			The values are normalized with the volume averaged kinetic energy of the original field.
			Note that these snapshots were taken after synchronization had taken place, so $\uu_1 + \ww$ is equal within numerical precision to the original field $\uu_2$.
			The latter cannot be distinguished by eye from the reconstructed field.  
			\label{fig:coarse versus fine}}
		\end{figure}

		Fig.~\ref{fig:error decay} also shows that the exponential decay rate $a$ becomes greater at larger $k_c,$ a natural result since $\ww$ then 
		lives on smaller and hence faster scales. We have studied this effect quantitatively. 
		The linear part of the trends in Fig.~\ref{fig:error decay} can be computed from the data by least-square error fitting  $a t + b$ 
		to the measured $\log_{10} \epsilon(t)$ in the region where the error is larger than the roundoff error floor, i.e. for $\epsilon$ between $1$ and $10^{-5}$.
		The behavior of the measured $a$ as function of $k_c$ depended on the various parameters of the simulations.
		To attempt to collapse the results, various non-dimensionalizations for $a$ and $k_c$ were tested.
		It was found that good collapse is observed when using Kolmogorov (viscous) scales for both the cutoff wavenumber as well as the synchronization exponent, i.e. to plot  $a \tau_K$ versus $k_c \eta_K$.
		See Fig.~\ref{fig:exponent stats}.
		To document the scatter due to possible lack of statistical convergence, the duration of ``exponential decay'' was split in half for each individual run, the corresponding pair $(k_c \eta, a \tau_K)$ was computed for each of the resulting $\epsilon(t)$ histories, and then the average over all the simulations with the same resolution (or Reynolds number) was computed.
		These are the results that are presented as symbols in Fig.~\ref{fig:exponent stats}.
		Error bars are for maximum and minimum values. The results collapse reasonably well, 
		although the lines seem to shift a little to the right with increasing resolutions. This hints at a slight Reynolds number dependence, 
		which is expected due to intermittency \cite{paladin_vulpiani_1987,yakhot_sreeni_2005}.
	         The results of Fig.~\ref{fig:exponent stats} are parameterized well by a linear fit $a\tau_K \approx - \beta(k_c\eta_K-0.15)$, implying that 
	         synchronization of small scales to large scales occurs only if the cutoff wavenumber is such that $k_c \eta_K > 0.15$.
                   Using the correspondence $r_c = \frac{\pi}{k_c}$, 
		this denotes scales smaller than $r_c <20 \eta_K$, i.e. in the transition zone between the inertial and viscous ranges.

		The key point to be taken from this study is that it is possible to reconstruct perfectly the small scales of a turbulent flow from coarse-grained data.
		If the velocity of a turbulent fluid is sampled on a spatial grid even as coarse as 10-15 times the Kolmogorov scale, these (time-dependent) data 
		can be refined to their original resolution, in the sense that the subdynamics of small scales after a suitable time will synchronize with the large-scale dynamics. 
		Figure \ref{fig:coarse versus fine} shows what this refinement implies: fine details of small-scale structures that are smeared out in the coarse-grained field 
		reappear, as if by magic, when refined by computing the subdynamics. Of course, synchronization takes time.
	          For example, assuming that $k_c \eta_K \approx 1/4$ as in Fig.~\ref{fig:coarse versus fine}, and assuming that a precision of $\epsilon = 10^{-3}$ is  desired, an interval of about $50 \tau_K$ is needed.
		This translates into about $65 / R_\lambda$ in units of the integral time, significantly  less then an integral time for moderately large values of $R_\lambda$.
		Doubling of this interval would lead to an error of order $10^{-6},$ at the lower limit for single precision computations.

Our results offer some support to the current practice of DNS with grid spacing of order $\eta_K,$ since they suggest that there may 
be an exact solution of 3D NS which, when coarse-grained to the grid scale, agrees with the finite-resolution simulation. Tiny scales 
much smaller than $\eta_K$ may be present, but completely slaved to the super-Kolmogorov scales. To more fully address these issues,
numerical experiments on CS must be performed with approximate data for $\uu_1$ which come not from a projection of a fine-grained 
solution $\uu_2$ but instead from a pseudospectral DNS with cutoff wavenumber $k_c.$ Outstanding issues are whether CS will 
occur for such approximate $\uu_1$ and whether the reconstructed field $\uu_2=\uu_1+\ww$ is then a solution of the fine-grained 
equations.  These questions are currently under active investigation. The size of the smallest length-scale in turbulence is of interest 
not only for physical theory but also for fundamental mathematical theory of 3D incompressible NS.  The Clay Millenium Prize problem 
on that equation concerns whether its solutions at sufficiently high Reynolds numbers may develop actual singularities, with velocities 
exploding to infinity at the singularity and smallest length scale going to {\it zero} \cite{fefferman_2006}. In nature, physical effects beyond 
incompressible NS would cut off the singularity at some tiny length-scale, but the observable manifestations should be striking. There is 
presently no empirical evidence whatsoever for such ``Leray singularities'', but this may be due to limited resolution or statistics of current 
numerical and experimental studies.  If such singularities occur anywhere at all, high Reynolds turbulent solutions are perhaps the most likely 
venue. Better understanding  of the interactions between inertial range and far dissipation range modes in turbulent NS flows should help to illuminate this problem.  

		\subsection{Acknowledgments}
		This work is supported by the National Science Foundation's CDI-II program, project CMMI-0941530, with additional support through grant NSF-OCI-108849.
		
%

\end{document}